\documentclass[prb,twocolumn,showpacs]{revtex4}   
\usepackage{graphicx}

\begin{document}

\title
{
Quantum Phase Transitions in the One-Dimensional 
$S=1$ Spin-Orbital Model: Implications for Cubic Vanadates 
}

\author{Satoshi Miyashita}
\author{Akira Kawaguchi}%
 \thanks{Present address: NTT Basic Research Laboratories,  Atsugi, 
Kanagawa 243-0198, Japan.}
\author{Norio Kawakami}
\affiliation{%
Department of Applied Physics, Osaka University, 
Suita, Osaka 565-0871, Japan}%

\author{Giniyat Khaliullin}
\affiliation{
Max-Planck-Institut f\"ur Festk\"orperforschung,
Heisenbergstrasse 1, D-70569 Stuttgart, Germany}%

\date{\today}

\begin{abstract}
We investigate ground-state properties and quantum phase transitions 
in the one-dimensional $S=1$ spin-orbital model relevant to
cubic vanadates. Using the density matrix renormalization group, 
we compute the ground-state energy, the magnetization 
and the correlation functions for different values of
the Hund's coupling $J_H$ and the external magnetic field. 
It is found that the magnetization jumps at a certain critical field, 
which is a hallmark of the field-induced first-order phase transition.
The phase transition driven by $J_H$ is also
of first order. We also consider how the lattice-induced 
ferro-type interaction between orbitals modifies the phase diagram, 
and discuss the results in a context of the first-order
phase transition observed in YVO$_3$ at 77K.
\end{abstract}
\maketitle

\section{Introduction}
Orbital fluctuations in correlated electron systems 
have attracted much interest recently. 
They play a particularly important role in transition metal oxides, 
where the low-energy physics is controlled by 
the orbital as well as spin degrees of freedom. 
Orbitals of $d$-electrons possess 
characteristic  spatial anisotropy, which sometimes results 
in remarkable phenomena such as the reduction of effective dimensions.
For instance, in cubic vanadates such as  YVO$_3$ and 
LaVO$_3$, an effective one-dimensional (1D) system is formed along the 
$c$-axis at low temperatures, which has been shown by 
a number of experiments 
\cite{Kawano,Ren-N,Ren-PRB,Noguchi,Miyasaka,Blake,Ulrich} 
on spin and orbital orderings. For YVO$_3$, there are two magnetic phases: 
$C$-type antiferromagnetic order 
(ferromagnetic along $c$-axis and antiferromagnetic in $ab$ plane) 
at higher temperatures $77K < T <114K$, and $G$-type spin order  
(antiferromagnetic in all three directions) 
at lower temperatures $T < 77K$.\cite{Blake} 
As for the orbital sector, 
the $G$-type orbital-antiferromagnetic correlations develop  
below $T\sim200K$ and it changes to the $C$-type 
order (in-phase alignment of orbitals along the $c$-axis) 
below $T=77K$. On the other hand, in LaVO$_3$, the magnetic order is 
always of the $C$-type and the orbital correlations are 
of the $G$-type.\cite{Miyasaka}  
On empirical grounds, difference in the ground state of these 
two compounds is usually attributed to larger lattice distortions in 
YVO$_3$.\cite{Blake} More specifically, an increase 
in octahedral tilting in YVO$_3$ 
supports ferro-type alignment of orbitals along the $c$-axis, hence 
stabilizing the $G$-type spin order in the ground state of YVO$_3$. This
argument has been supported by electronic structure 
calculations.\cite{Mizokawa2} In vanadates, there are two 
electrons in the threefold degenerate $t_{2g}$ level, 
and it is believed that orbital degrees of freedom are crucial 
for understanding the interplay between different orderings 
in these compounds and their highly unusual magnetic 
properties.\cite{Ren-N,Ren-PRB,Noguchi,Miyasaka,Blake,Ulrich} 

Based on the above experiments, an effective 
spin-orbital model has been proposed,\cite{Khaliullin} which may 
properly describe low-energy physics of the cubic vanadates. 
The 1D version of this model consists of the $S=1$ Heisenberg spin model 
with two-fold degenerate orbitals 
[see Eqs.(\ref{so-model}) and (\ref{operator})]. 
It has been demonstrated that 
this effective 1D model contains rich and essential properties
inherent in spin-orbital systems.\cite{Sirker}
A similar quasi-1D orbital model has been also
proposed for LaVO$_3$.\cite{Motome}

In this paper, 
by calculating the ground-state energy, the spin and orbital
correlation functions, and the magnetization curve
by means of the density matrix renormalization group (DMRG) method,
\cite{DMRG} we investigate quantum phase transitions of
the 1D $S=1$ spin-orbital model, which may be relevant for explaining
the magnetic properties of cubic vanadates, in particular, for 
understanding the phase transitions observed in YVO$_3$.
We obtain the phase diagram in the plane 
of the Hund's coupling $J_H$ and the external magnetic field $h$.
To make the model more realistic, 
we introduce also an additional ferromagnetic interaction $V$ to
the orbital sector, which may drive the system 
to an orbital-ferromagnetic phase.
We discuss the importance of this interaction for the
low-temperature phase of YVO$_3$ below 77K.

This paper is organized as follows. After brief discussion of the
model in Sec. II, we present in Sec. III the DMRG results
for the ground-state energy,
the spin/orbital correlation functions in a magnetic field, and 
show the magnetic phase diagram at absolute zero. In Sec. IV, we 
address the effects of the ferromagnetic orbital coupling driven by lattice
distortions.  Brief summary is given in Sec. V.

\section{Model}

We start with 1D version of an effective Hamiltonian, which has been
proposed to describe the spin and orbital properties
of cubic vanadates: \cite{Khaliullin,Sirker} 
\begin{eqnarray}
 {\cal H} &=& 
J \sum_{i} 
         [\frac{1}{2}
           ({\bf S}_{i} \cdot {\bf S}_{i+1} + 1)
           {\hat J}_{i,i+1} + {\hat K}_{i,i+1}
         ]
\nonumber \\
   &-& h \sum_{i} S_{i}^{z}, 
\label{so-model}
\end{eqnarray}
where ${\bf S}_{\it i}$ is the $S=1$ spin operator at the $i$-th
site. $J (\equiv 4t^2/U)$ is 
the superexchange interaction scale, which we will take as the
energy unit in the following discussions. 
${\hat J}_{i,j}$ and ${\hat K}_{i,j}$ are the operators acting
on the doubly-degenerate orbital degrees of freedom, and are given as
\begin{eqnarray}
 {\hat J}_{i,j} &=& ( 1+2R ) 
           \left( 
                  {\bf T}_{i} \cdot {\bf T}_{j} + \frac{1}{4}
           \right)
        - r 
           \left( 
                  T^z_{i} T^z_{j} + \frac{1}{4}
           \right)
        -R , 
\nonumber \\
 {\hat K}_{i,j} &=& R 
           \left( 
                  {\bf T}_{i} \cdot {\bf T}_{j} + \frac{1}{4} 
           \right)
        + r 
           \left( 
                  T^z_{i} T^z_{j} + \frac{1}{4} 
           \right), 
\label{operator}
\end{eqnarray}
where ${\bf T}_i$ is the $T=1/2$  pseudo-spin operator for the orbital
sector. The terms proportional to 
$R=\eta/(1-3\eta)$ and $r=\eta/(1+2\eta)$ with $\eta=J_H/U$ originate 
from the Hund's coupling (normalized by the on-site Coulomb
interaction $U$). The effective model (\ref{so-model})
is derived via the second-order 
perturbation calculation in $t/U$ ($t$ is the electron hopping amplitude) 
for the cubic vanadates,  
where two correlated electrons occupy $t_{2g}$ orbitals at each 
site. The microscopic derivation  can be 
found in Ref.~\onlinecite{Khaliullin}. Physically, the external field $h$
in the above Hamiltonian can be regarded as the interchain molecular field
present in a realistic 3D system. \cite{note1}  

For $h=\eta=0$, it has been  
concluded that the ground state of the model (\ref{so-model})
is the orbital valence bond (OVB) solid state for  which
both of the spin and orbital sectors are in 
a dimer-type singlet state with gapful excitations. 
\cite{Sirker,Shen,Ulrich} 
Furthermore, it was obtained \cite{Sirker,Ulrich}
that a first-order transition 
from the OVB solid state to the fully polarized spin state
with gapless orbital singlet 
may occur around a critical value $\eta_c\sim0.11$.
Since the analysis of Refs.~\onlinecite{Sirker,Ulrich} 
is based on the finite-temperature 
DMRG method, it is difficult to discuss the 
behavior at lower temperatures. It is thus 
desirable to study zero-temperature  properties
to establish the precise  phase diagram, which is 
addressed below in this paper.  We further investigate 
the effects of a magnetic field, as well as of an additional ferromagnetic 
orbital coupling (which will be specified in Sec. IV). 

Before closing this section,
we make a brief comment on the spin-orbital model.
The $S=1$ model (\ref{so-model}) may be regarded as a specific
extension of the $S=1/2$ spin-orbital model, which 
has been studied extensively.
\cite{Suther,Affleck,Itakura,Azaria,Itoi,Pati,Yamashita} 
The ground-state phase diagram of the latter model has been
already established by the DMRG method.
\cite{Itoi,Pati,Yamashita} A simple isotropic
$S=1$ generalization leads to the model which has the 
OVB ground state.\cite{Sirker,Shen}
We will see below that the extended model, which includes anisotropic
orbital interactions, 
the magnetic field and also an additional ferromagnetic orbital interaction, 
exhibits interesting quantum phase transitions.

\section{Magnetic Phase Diagram}

\begin{figure}[htb]
\begin{center}
\includegraphics[width=10cm]{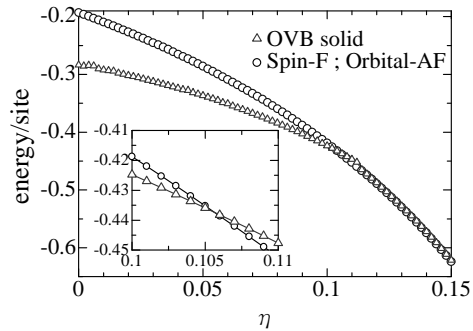}
\end{center}
\vskip -20mm
\caption{
The energy per site calculated by the infinite DMRG method
as a function of the normalized Hund's coupling $\eta$:
the OVB solid state (triangle) and the spin-ferromagnetic state (circle).
The inset shows the detail of the $\eta$-dependence
in the vicinity of $\eta_c$. 
}
\label{ene-h0}
\end{figure}
\begin{figure}[htb]
\begin{center}
\includegraphics[width=10cm]{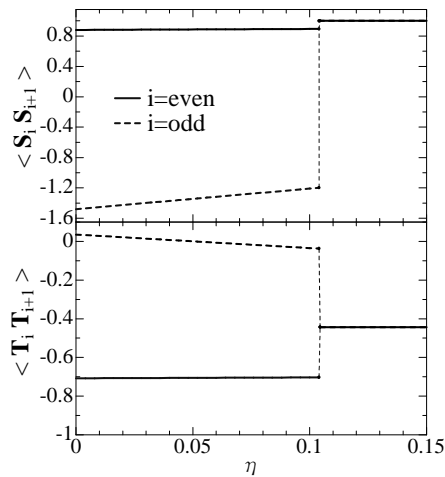}
\end{center}
\vskip -7mm
\caption{
Spin (upper panel) and orbital (lower panel) correlation functions
as a function of the Hund's coupling $\eta$.  
The site index $i$ is even (solid line) or odd (dashed line). 
The even-odd $i$-dependence reflects a dimer property of
the OVB solid state. 
}
\label{corre-h0}
\end{figure}

We first consider the effect of the Hund's coupling.  To this end, 
we precisely determine the critical value $\eta_c$ between the 
OVB solid state and the spin-ferromagnetic state.
We show the $\eta$-dependence of the energy and the 
correlation functions calculated by the 
infinite DMRG method \cite{DMRG} 
for $h=0$ in Figs.~\ref{ene-h0} and \ref{corre-h0}. 
It is seen from Fig.~\ref{ene-h0} that
for small values of the Hund's coupling, the energy of the OVB solid state 
is lower than that of the spin-ferromagnetic state. 
In this region, both of the spin and orbital
correlation functions show the 
even-odd $i$-dependence reflecting a dimer character of the
OVB solid state (see Fig.~\ref{corre-h0}).
More precisely, for $i=even$, 
two adjacent orbitals form  nearly perfect dimer pairs 
(namely, $\langle {\bf T}_i \cdot {\bf T}_{i+1} \rangle$ 
is close to $-0.75$) while the spins prefer the ferromagnetic alignment
with a positive value of  
$\langle {\bf S}_i \cdot {\bf S}_{i+1} \rangle \simeq 0.9$. 
On the other hand, for $i=odd$, 
the value of the orbital correlation is almost zero 
while  spins form a singlet pair with
$\langle {\bf S}_i \cdot {\bf S}_{i+1} \rangle \simeq -1.5$, which is 
much less than the value ($=-2$) expected for a spin-1 
isolated dimer. \cite{note2} 
The above behavior is characteristic of the OVB solid state. 
Increasing $\eta$, the energy of the OVB solid state is getting closer to 
that of the spin-ferromagnetic state, 
and eventually a first-order phase transition occurs between
these two phases at a  critical value of
$\eta_c = 0.1055 \pm 0.0005$.  The detail around the 
transition is shown in the inset of Fig.~\ref{ene-h0}. 
Beyond the critical value $\eta_c$,  the even-odd $i$-dependence of 
the correlation functions disappears and 
both of the spin and orbital correlations become 
spatially uniform with 
\begin{eqnarray}
\langle S^z_iS^z_{i+1}\rangle = 1, \hskip 5mm
\langle {\bf T}_i \cdot {\bf T}_{i+1}\rangle = -\ln2+\frac{1}{4},
\label{corr}
\end{eqnarray}
as should be expected.\cite{bethe}
Our conclusion that the ground state changes 
from the OVB solid 
to the spin-ferromagnetic state, 
and also the value of $\eta_c$ at which the transition occurs, 
is in good agreement with 
Ref.~\onlinecite{Sirker}.

The effective spin (orbital) exchange coupling constants, which are determined
by local orbital (spin) correlations, change drastically
at the phase transition. Just below $\eta_c$, the spin interactions  
$J_s=\langle {\hat J}_{i,i+1}\rangle /2$ are equal to $J_F \simeq -0.38$ 
within the ferromagnetic bonds, and $J_{AF} \simeq 0.05$ between the spins
belonging to different dimers. Above the transition, $J_s \simeq -0.21$ in 
all the bonds, supporting uniform Heisenberg-like orbital dynamics.  
Similarly, the orbital exchange constant
$J_{orb}=(1+2R)\langle {\bf S}_{i} \cdot {\bf S}_{i+1}+1 \rangle /2+R$ 
changes from: $J_{orb} \simeq 1.4$ ($\simeq 0.02$) in 
strong (weak) bonds below 
$\eta_c$, to the uniform value $J_{orb}\simeq 1.46$ above $\eta_c$.
  
\begin{figure}[htb]
\begin{center}
\includegraphics[width=10cm]{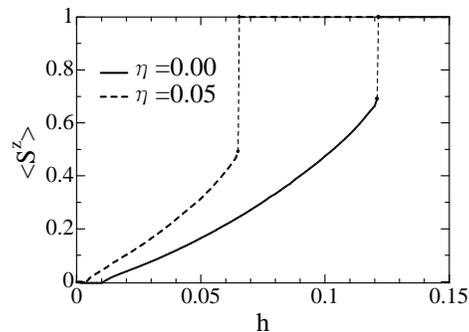}
\end{center}
\vskip -22mm
\caption{
Magnetization curves for 
$\eta=0$ (solid line) and $\eta=0.05$ (dashed line). 
The magnetization becomes finite for $h>h_{c1}$ 
and exhibits a jump at $h_{c2}$ characteristic of
the first-order phase transition. 
}
\label{mag}
\end{figure}

Let us next consider the magnetic properties of the  
model (\ref{so-model}) for finite values of $h$. 
We have calculated the magnetization curve 
as well as the correlation functions. The results of the
magnetization are shown in Fig. \ref{mag}  for 
two typical choices of the Hund's coupling.
As seen from the case of $\eta=0$ in Fig.~\ref{mag}, 
the ground state in small fields is still in the OVB solid state,
 rendering the magnetization  zero up to a critical field $h_{c1}$
due to the existence of the spin gap. Beyond the 
critical field $h_{c1}$, the spin sector becomes 
gapless, resulting in the square-root increase 
of the magnetization near the critical
field $h_{c1}$.  We note that this transition in the 
spin sector is a kind of Pokrovsky-Talapov transition, i.e.
an insulator to gapless-liquid transition triggered by 
filling-control.\cite{prokovsky}
Therefore, the ferro/antiferromagnetic alternating spin 
correlation still exists even in the gapless phase. The 
magnetization then smoothly increases until we encounter a
sudden jump  at the second  critical field $h_{c2}$, 
where the system undergoes a first-order
transition  to the fully polarized spin
state with the orbital sector forming a gapless $T=1/2$ Heisenberg 
pseudo-spin chain. Note that in contrast to the spin sector,
the orbital sector is always gapful for $0 \leq h < h_{c2}$.
For $\eta=0$ ($\eta=0.05$), the corresponding
critical field is  $h_{c2}=0.122$ ($h_{c2}=0.065$). 

\begin{figure}[htb]
\begin{center}
\includegraphics[width=10cm]{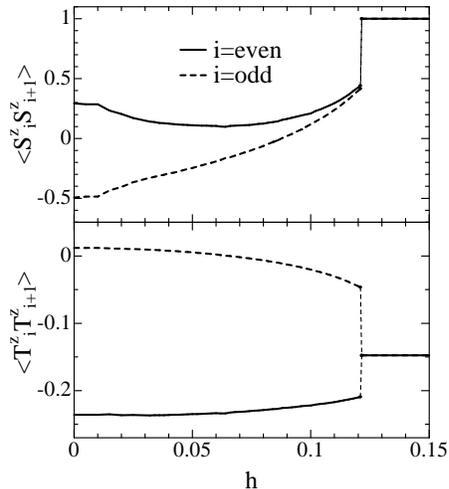}
\end{center}
\caption{
Spin (upper panel) and orbital (lower panel) correlation functions 
for $\eta=0$ in a  magnetic field.
The site index $i$ is even (solid line) or odd (dashed line).
}
\label{corre-eta0}
\end{figure}

Shown in Fig. \ref{corre-eta0} are the correlation functions 
for $\eta=0$ under magnetic fields. As mentioned above,
in weak fields,  both of 
spin and orbital correlations exhibit the even-odd dependence 
characteristic of the OVB solid state. Their values are unchanged up
to $h_{c1}$, beyond which the even-odd dependence 
is gradually suppressed.  Notice that the spin correlation function 
shows a cusp singularity at $h_{c1}$ while the orbital correlation function 
smoothly changes. This is because the 
spin sector shows an insulator to gapless-liquid transition, whereas the 
orbital sector still stays in the gapful phase, which is 
affected by magnetic fields indirectly via the spin-orbital
interaction.  In magnetic fields beyond $h_{c2}$, 
the spin and orbital correlation functions take the values
given in (\ref{corr}) corresponding to 
the fully polarized spin state with a gapless orbital Heisenberg chain. 
Similar behavior in the field dependence of the correlation
functions is observed for other choices of the Hund's coupling,
as far as the ground state is in the OVB solid state at $h=0$.

\begin{figure}[htb]
\begin{center}
\includegraphics[width=10cm]{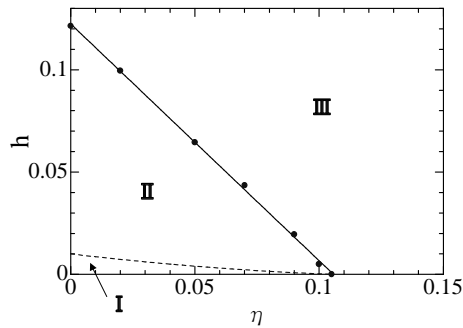}
\end{center}
\vskip -22mm
\caption{
Magnetic phase diagram:
Phase I represents the OVB solid state with 
zero  magnetization (spin-dimer state). 
Phase I$\!$I is also in the OVB solid state, but with
a finite magnetization $0\leq\langle S^z_i \rangle \leq1$.
Phase I$\!$I$\!$I is the spin-ferromagnetic phase with a
gapless $T=1/2$ orbital chain.
The first-order phase transition occurs at the 
boundary denoted by the solid line. 
}
\label{eta-h-phase}
\end{figure}

We thus end up with the zero-temperature magnetic phase diagram, 
which is shown in Fig. \ref{eta-h-phase}. 
There are three distinct phases in the $\eta$-$h$ plane. 
In the region of weak fields and weak Hund's couplings,
the OVB solid state is stabilized in the  phases I and I$\!$I;
the orbital sector is gapful while 
the spin sector is gapful (gapless) in the phase I (I$\!$I). 
The transition from the phase I to I$\!$I is of Pokrovsky-Talapov
type. On the other hand, the transition from the phase I$\!$I 
to the spin-ferromagnetic phase I$\!$I$\!$I is of first order.

For the spin/orbital dimerized state at $\eta=0$, 
we estimated the spin gap from the magnetization curve and obtained 
$\Delta_s\sim0.01J$. This is actually much larger than the Haldane gap 
that one would expect from a simple picture of spin-2 dimers coupled 
by a weak antiferromagnetic exchange. Indeed, 
at $\eta=0$ an effective spin exchange constant 
between the dimers is estimated as 
$J_{eff}=\frac{1}{8}\langle {\bf T}_{i} \cdot {\bf T}_{j} + 
\frac{1}{4} \rangle \simeq 0.036$. This would give the spin-2 Haldane gap  
about $0.085 J_{eff}\simeq 0.003$ only.\cite{Scholl,Yamamoto} 
The large spin gap in the spin-orbital model is due to a dynamical 
coupling between two sectors, as discussed in Ref.~\onlinecite{Sirker}.    
We notice that our result for the spin gap is still much smaller than  
$\Delta_s\sim0.041J$, obtained in Ref.~\onlinecite{Sirker} 
from the finite-temperature spin susceptibility $\chi$ 
computed by the transfer-matrix DMRG method. 
As the latter method is available at finite temperatures only, 
Ref.~\onlinecite{Sirker} estimates the spin gap by extrapolating the data
with a function of $\chi \sim \exp (-\Delta_s/T)/\sqrt{T}$. We think that
the above discrepancy may arise from this (somewhat ambiguous)
extrapolation procedure. 

\section{Effects of the Ferromagnetic Orbital Interaction}

For representative values of $J_H\simeq0.68$eV and 
$U-\frac{20}{9}J_H \simeq 4.5$eV for vanadates, \cite{Mizokawa}  
one obtains that the ratio $\eta=J_H/U\simeq 0.11$ is slightly above its
critical value. Therefore, our results are consistent with the 
experimental fact that the ferromagnetic chain is stabilized along the 
$c$-axis in the ground state of LaVO$_3$.\cite{Miyasaka}
However, the superexchange model (\ref{so-model}) alone is not sufficient
to understand the origin of the spin-$G$ type ground state in YVO$_3$, and
also to understand the physics behind the spin/orbital reordering at $T=77K$.
In order to describe low-temperature 
properties of YVO$_3$ more precisely, it is desirable to take into account
the pure-orbital interaction in addition to the spin/orbital
couplings involved in the Hamiltonian (\ref{so-model}).
The orbital interaction to be included has 
the following Ising-like form,\cite{Khaliullin}
\begin{eqnarray}
 {\cal H}_{o} = - V \sum_i T^z_{i} T^z_{i+1}. 
\label{o-ising}
\end{eqnarray}
This describes a ferromagnetic orbital coupling, which competes with 
the antiferromagnetic orbital coupling in the  Hamiltonian
(\ref{so-model}). 
Physically, this term is important 
because apart from the superexchange mechanism there is also
a lattice contribution to the orbital interactions. As 
mentioned in the introduction, tilting of octahedra in perovskites
(caused by ionic mismatch effects) leads to 
a tendency that orbitals are aligned ferromagnetically
along the $c$-axis. This effect  
is stronger when we go from La- to Y-based compound.\cite{Mizokawa2} 
We thus expect that the ferromagnetic Ising orbital coupling 
along the $c$-axis may be particularly relevant for YVO$_3$, 
in order to explain why its ground state shows 
spin-antiferro/orbital-ferromagnetic structure 
along the $c$-axis. 

\begin{figure}[htb]
\begin{center}
\leftskip -7mm
\includegraphics[width=10cm]{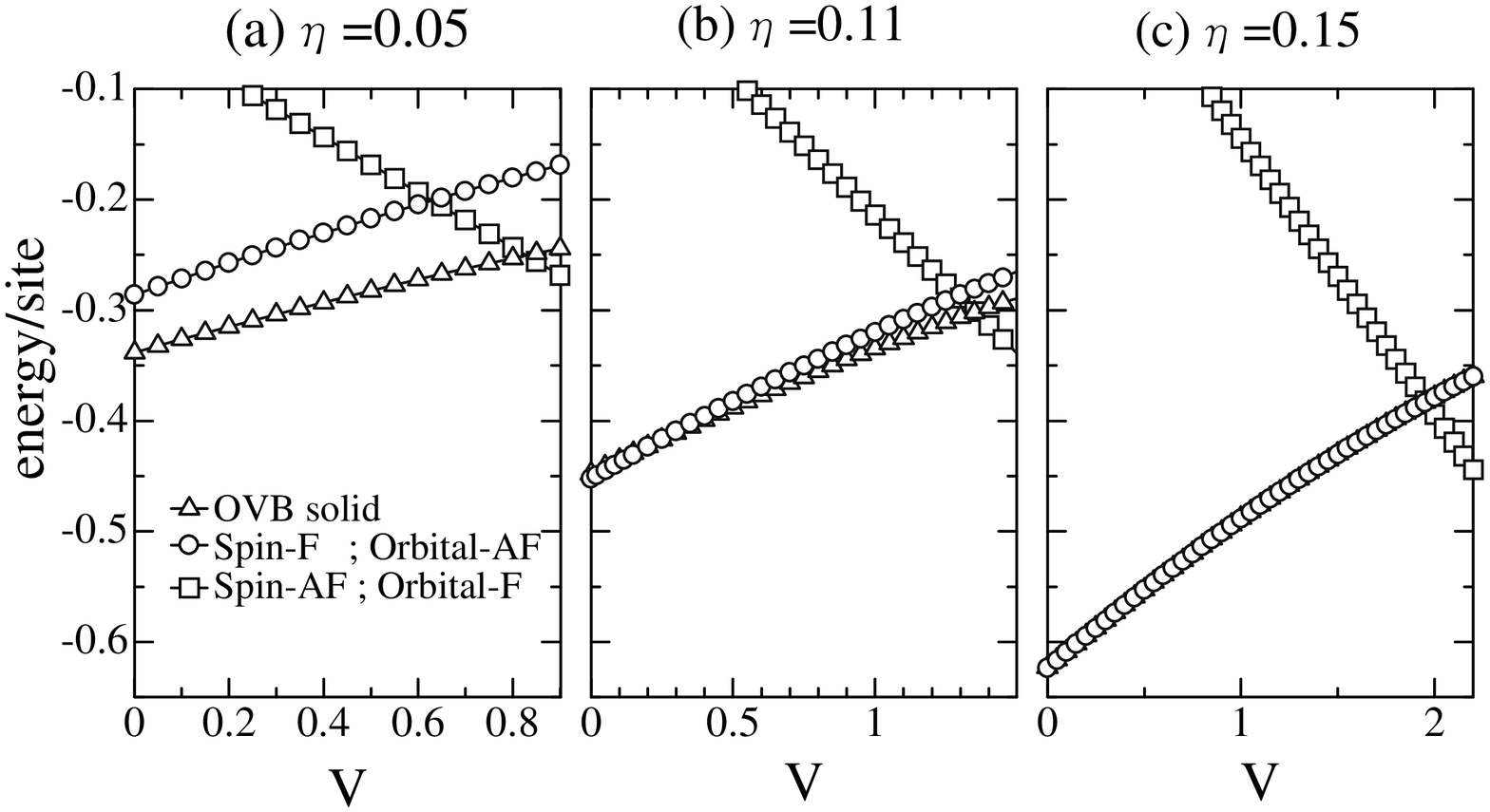}
\end{center}
\vskip -22mm
\caption{
The energy per site as a function of the orbital coupling $V$ for
several choices of $\eta$; the OVB solid state (triangle), 
the spin-ferromagnetic state (circle), orbital-ferromagnetic 
state (square). Notice that the first-order transition 
occurs once for (a) and (c), while it occurs twice for (b) when
the value of $V$ is increased.
}
\label{V-ene}
\end{figure}

We choose here three different values of the Hund's coupling
and calculate the energy as a function of V. 
In Fig.~\ref{V-ene}(a), the $V$-dependence of 
the energy for three competing states is
shown for the OVB ground state ($\eta=0.05$). There is a direct
first-order transition from the OVB state to the orbital-ferromagnetic
state. On the other hand, when the 
Hund's coupling is slightly above the critical 
value $\eta_c$ ($\eta=0.11$), we encounter different behavior,
as seen in  Fig.~\ref{V-ene}(b).  Namely, 
the ground state is the spin-ferromagnetic (orbital-ferromagnetic)
state for small $V$ (large $V$), but for intermediate $V$,
the OVB solid state is stabilized, which results from 
the competition of 
the above two (spin-ferromagnetic or orbital ferromagnetic) interactions. 
Stabilization of the {\it orbital disordered} 
OVB state by finite $V$-interaction (induced by {\it lattice distortion}) 
is a remarkable result. The physics behind this observation is that 
the lattice-driven interaction (\ref{o-ising}) introduces frustration 
into the orbital sector, competing with antiferro-type alignment of orbitals
due to the superexchange process. For even larger 
$\eta$ [Fig.~\ref{V-ene}(c)], $V$ induces a direct phase transition 
from the spin-ferromagnetic state to the orbital-ferromagnetic state.

\begin{figure}[htb]
\begin{center}
\includegraphics[width=10cm]{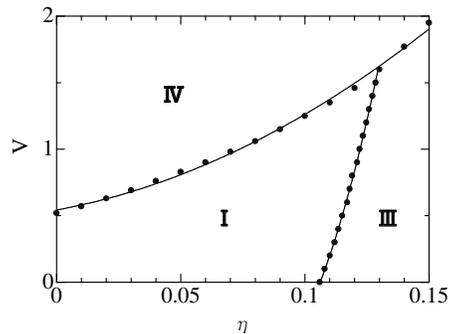}
\end{center}
\vskip -22mm
\caption{
Phase diagram in the $\eta$-V plane; 
Phase I is the OVB phase, I$\!$I$\!$I is  the 
spin-ferromagnetic phase, and 
I$\!$V is the orbital-ferromagnetic  phase 
stabilized by the orbital ferromagnetic Ising term. 
All the phase transitions are  of first-order. 
Note that the orbital (spin) sector in the phase
I$\!$I$\!$I (I$\!$V) is  a gapless-orbital (gapful-spin) liquid.
}
\label{eta-V-phase}
\end{figure}

The phase diagram thus obtained in the $\eta$-V plane is
shown in Fig.~\ref{eta-V-phase}. 
There are three distinct phases in this figure. 
For small $\eta$ and $V$, 
the OVB solid state is stabilized in the phase I, which
is driven to the orbital-ferromagnetic phase I$\!$V with the increase of 
$V$. In the latter phase I$\!$V, the spin sector is described
by an $S=1$ isotropic antiferromagnetic spin chain, and thus its
ground state is a spin liquid  with Haldane-gap
excitations. In the spin-ferromagnetic phase I$\!$I$\!$I, 
the orbital sector is described by the $XXZ$ pseudo-spin model,
which results from (\ref{so-model}) and (\ref{o-ising}):
\begin{eqnarray}
{\cal H}=\widetilde J \sum_{i}
{\bf T}_{i} \cdot {\bf T}_{i+1} - V \sum_{i}T^z_{i} T^z_{i+1}.
\label{effective}
\end{eqnarray}
Here $\widetilde J=J/(1-3\eta)$,
and we have discarded irrelevant constant terms.
According to the well known results for the $S=1/2$ $XXZ$ spin 
chain with ferromagnetic Ising anisotropy,\cite{xxz}
the ground state is in a gapless liquid phase for $V<2\widetilde J$,
and in a fully polarized ferromagnetic phase for $V \geq 2\widetilde J$.
Therefore it might be possible to
have the orbital-ferro/spin-ferromagnetic ground state 
in the region $V \geq 2/(1-3\eta)$ in units of $J$. However,
it turns out that 
since the first-order transition from the phase I$\!$I$\!$I to
I$\!$V occurs at the phase boundary shown in 
Fig.~\ref{eta-V-phase}, there is no available window 
of the parameters which can stabilize
the orbital-ferro/spin-ferromagnetic ground state.

The ferromagnetic orbital interaction $V$ may be relevant to the 
first-order phase  transition observed 
in YVO$_3$ at $T=77K$, where the 
spin ordering changes from the $G$-type (below $77K$) to the 
$C$-type (above $77K$) structure.
If we focus on the pattern of ordering in the 
$c$-direction, the transition occurs from the 
spin-antiferro/orbital-ferromagnetic ground state to 
the spin-ferro/orbital-antiferromagnetic state.
Furthermore, recent neutron scattering experiments have revealed 
a substantial modulation of the ferromagnetic spin 
couplings along the $c$-direction at $T>77K$, which indicates 
the appearance of orbital dimer correlations 
at finite temperatures.\cite{Ulrich}
According to the results obtained here, such a
temperature-driven first-order transition
may possibly occur when the system is located in the phase 
I$\!$V near the tricritical point in the phase diagram
(Fig.~\ref{eta-V-phase}), and thus close to 
the phases I$\!$I$\!$I and I. 
For example, $V\simeq 1.5$ may be a reasonable choice for $\eta\simeq0.12$,
which may give rise to the competition of the phases 
at finite temperatures. This is in fact quite realistic situation,
as the atomic value of Hund's coupling parameter 
for vanadates is indeed close to this value. \cite{Mizokawa}
Finally, we wish to mention that the 
gapless orbital-liquid (gapful spin-liquid)
state in the phase I$\!$I$\!$I (I$\!$V) is expected to 
show an orbital (spin) antiferromagnetic order when
realistic three-dimensional effects are properly taken into 
account, being  consistent with experimental findings.

\section{Summary}

We have investigated quantum phase transitions of the 
1D $S=1$ spin-orbital model relevant to cubic vanadates.
By means of the DMRG calculation,
we have determined the  zero-temperature phase diagram in a magnetic
field. We have found  that the magnetic field induces the 
phase transitions twice for the case of OVB ground state. 
The lower-field transition is of Pokrovsky-Talapov type, which 
drives the spin sector to a gapless spin liquid while keeping 
the orbital sector  always gapful.
On the other hand,  the high-field
transition is first order to the spin-ferromagnetic state, where the 
orbital state is driven to a gapless isotropic $T=1/2$ orbital chain.

We have considered the evolution of spin and orbital correlations 
as a function of the Hund's coupling, and obtained that the ground 
state of the superexchange model is spin-ferromagnetic for the
realistic values of $J_H$. This is consistent with the formation 
of ferromagnetic spin chains along the $c$-axis in LaVO$_3$. 

The introduction of the ferromagnetic Ising coupling $V$ 
to the orbital sector induces a first-order 
phase transition to the orbital-ferro/spin-antiferromagnetic state. 
This may explain the stabilization of the $G$-type spin order 
observed in YVO$_3$ at temperatures below 77K. We found also that
the intermediate values of $V$-interaction may stabilize 
the spin/orbital dimer phase. This phenomenon reflects competition 
between the superexchange and lattice effects, and, surprisingly enough,
it occurs for $J_H$-values realistic for vanadates. The proximity 
of different spin/orbital states in the ground state may result 
in an entropy-driven phase transition at finite temperature, 
similar to that observed in YVO$_3$.   
In the future study, it is desirable to discuss whether such a 
temperature-driven first-order transition may be indeed possible 
in the spin-orbital model including the ferromagnetic orbital
interaction and also the interchain coupling effects. 
This may provide a key to clarify some essential spin/orbital
properties of YVO$_3$ at low temperatures.

\section*{Acknowledgements}
This work was partly supported by a Grant-in-Aid from the Ministry 
of Education, Science, Sports and Culture of Japan. 
A part of computations was done at the Supercomputer Center at the 
Institute for Solid State Physics, University of Tokyo
and Yukawa Institute Computer Facility. 

%



\begin{thebibliography}{99}
\bibitem{note1}
We do dot consider here the interchain interactions in orbital sector. 
As discussed in Refs.~\onlinecite{Khaliullin,Motome}, orbital interactions
in the spin C-phase are essentially of 1D nature.
\end{thebibliography}
\end{document}